\documentclass[aps,amssymb,article,pra,twocolumn,showpacs,floatfix]{revtex4}

\usepackage{epsf}
\usepackage{amsmath}
\usepackage{graphicx}
\begin{document}
         
\title{X-ray and neutron emissions by shock waves}

\author{Boris I. Ivlev}

\affiliation{Instituto de F\'{\i}sica, Universidad Aut\'onoma de San Luis Potos\'{\i},\\ 
San Luis Potos\'{\i}, 78000 Mexico}

\begin{abstract}

Experimentally observed X-ray and neutron emissions by acoustic perturbations of liquids and solids look paradoxical. All acoustically driven effects are extremely adiabatic with respect to
typical times $\hbar/1keV\sim 10^{-18}s$ for X-ray and $\hbar/1MeV\sim 10^{-21}s$ for neutron processes. A direct application of these mechanism would result in negligible (exponentially small)
emission probabilities. As argued in this paper, high energy processes of X-ray and neutron emissions are caused by electron transitions in deep ($\sim 1MeV$) and narrow ($10^{-11}cm$) anomalous 
wells created by the local reduction of electromagnetic zero point energy. The formation of anomalous states cannot be described solely by quantum electrodynamics since the mechanism of electron 
mass generation is involved. 

\end{abstract} \vskip 1.0cm


%\keywords{X-ray, neutrons, Higgs field, shock waves}

\maketitle

\section{INTRODUCTION}
\label{intr}
In experiments \cite{KOR1,KOR2} emission of X-rays in the $keV$ range was observed from a matter acted by shock waves. In \cite{KOR1} the iron sample, emitted X-rays, was acted by shock waves 
produced outside it. In Ref.~\cite{KOR2} the copper medium was involved. The X-ray emission under shock waves has been also indicated in Ref.~\cite{HAG}. 

In a liquid or a solid electrons are adiabatically dragged by the shock wave with its velocity $v\sim 10^{5}cm/s$. Therefore a shock wave itself cannot result in Bremsstrahlung of the $keV$ quanta. 

A ``jolt'' of atoms by the shock front, of the width $l$ of a few Angstroms, occurs during $l/v\sim 10^{-13}s$ which is extremely adiabatic with respect to the short time $\hbar/1keV\sim 10^{-18}s$. 
Therefore atoms excitation, which may result in a characteristic emission in the $keV$ region, is also impossible.  

In experiments \cite{KOR1} the metallic surface, acted by shock waves and emitted X-rays, was far from the liquid producing cavitation mediated shock waves. So the internal dynamics of bubbles in 
the cavitating liquid was not relevant for the X-ray emission. 

In experiments \cite{CARD1,CARD2} neutron emission was registered from the acoustically driven cavitating water solution of iron salts. It was no residual radioactive isotopes in the matter since 
neutron emission appeared gradually upon increasing of acoustic power of the generator producing cavitation. As known, each cavitation bubble, of the micron size, rapidly shrinks as in the 
phenomenon of sonoluminescence \cite{PUTT,BRE}. During that implosion the energy is accumulated inside the bubble resulting in the UV emission up to $10eV$ of quanta energy \cite{PUTT,BRE}. This
energy is much smaller than the nuclear $MeV$ scale. Therefore the neutron source in \cite{CARD1,CARD2} was not inside micron size bubbles. 

Bubbles, created in a liquid by the acoustic source, emit cavitation mediated shock waves \cite{GOM,BREN,VOG}. But the ``jolts'' of atoms by those shocks are too adiabatic to produce neutron  
emission. 

In Refs.~\cite{CARD3,CARP1,CARP2} neutron emission from solids was reported. The driving force for that was shock waves generated by fractures of microdefects in iron-reach natural rocks. 
Generation and propagation of shock waves in elastoplastic solids is described on the level of solid state physics \cite{DUR}. These processes are extremely adiabatic with respect to the nuclear 
time $\hbar/1MeV\sim 10^{-21}s$ and they cannot lead to nuclear processes.

Therefore there is a common feature of X-rays and neutron emissions by shock waves. This is the paradoxical mismatch between the adiabatic acoustic perturbations, $10^{-13}s$, and the fast 
response, which is  $10^{-18}s$ for X-rays emission and $10^{-21}s$ for neutron emission. Directly calculated probability of those processes $\exp(-A)$ depends on the ratio $A$ of two characteristic 
times. For X-ray emission $A\sim 10^5$ and for neutron emission $A\sim 10^8$, that is equivalent to zero probability. $\exp(-A)$ is proportional to the high frequency fraction in the adiabatic 
perturbation. So it is unclear how acoustic (low energy) perturbations can cause high energy processes. Low energy nuclear reactions are impossible due to high Coulomb barriers to be passed by 
charged nuclei. 

It is argued in the paper that the high energy ($keV$ for X-rays and $MeV$ for neutrons) comes from anomalous electron wells where electrons go down in energy \cite{IVLEV3,IVLEV4}. These deep 
($\sim MeV$) and narrow ($\sim 10^{-11}cm$) wells are formed by the local reduction of zero point electromagnetic energy. Similar reduction, on much larger spatial scale, occurs in the Casimir 
effect. 

Anomalous wells correspond to the ground state. But the probability to form those state by usual electrons in condensed matter is of the type $\exp(-1000)$. The reason is smallness of the matrix 
element due to the difference in spatial scales of usual electrons and ones in anomalous wells. In contrast, under a perturbation of a short scale ($10^{-11}cm$), the probability of anomalous state 
creation is not small. Such perturbation is provided by rapidly varying in space charge density related to reflected shock waves. The formation of anomalous states cannot be described solely by 
quantum electrodynamics (QED) since the mechanism of electron mass generation is involved. 

So the neutron emission occurs by the usual (high energy) nuclear reactions. These reactions are caused by $MeV$ quanta generated by electron transitions in anomalous wells. It is unusual that the 
electron subsystem in condensed matter relates to $MeV$ energies which are typical for nuclear processes. 
\section{X-RAY EMISSION FROM A METAL ACTED BY SHOCK WAVES}
\label{kor}
\subsection{Experiments}
\label{korexp}
In this section we analyze the set of experiments on emission of X-ray beams of the energy in the $keV$ region \cite{KOR1,KOR2}. This emission from a steel plate ($3mm$ thick) is schematically 
shown in Fig.~\ref{fig1}. The high pressure ($600atm$) water jet of the diameter no more than $1mm$ was on the distance of approximately $3cm$ to the left from the plate in Fig.~\ref{fig1}. Next to 
he plate, separated by $1cm$ from it, a flat X-ray film was placed. 

An increase of pressure in the water jet leads to the bubble cavitation. Fast collapse of those bubbles results in shock waves generation \cite{GOM,BREN,VOG}. Shock waves propagate from the water 
jet and the metallic rode (Fig.~{\ref{fig1}}) in the air toward the metal plate. Inside the plate shock waves are generated as shown in Fig.~\ref{fig1} \cite{PRE,DUR}. The surface $A$ of the plate 
emits X-ray that are registered on the X-ray film in Fig.~\ref{fig1}. 

The X-ray spectrometer indicated the continuous X-ray spectrum in the range $(1-6)keV$. Regardless of spectrometer data, the $keV$ range of emitted X-rays was established in \cite{KOR1} independently.
Two X-ray films rolled tightly together in the form of a cylinder and placed around the water jet, shown the attenuation of exposure on the external film according to the attenuation length of
X-rays of 2 $keV$. 

X-ray emission was independently observed as images on the X-ray film \cite{HAG} in a configuration similar to \cite{KOR1}. 
\subsection{The paradox}
\label{par}
\begin{figure}
\includegraphics[width=6.5cm]{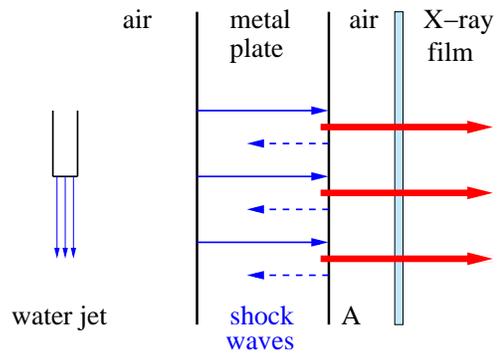}
\caption{\label{fig1} X-ray emission in experiments \cite{KOR1}. The high pressure water jet  escapes from the metallic rode and, due to cavitation, produces shock waves acted the metal plate. The 
thin solid arrow denotes the acoustic/shock wave propagating in the metal plate (of $3cm$ thickness) toward its surface $A$. The dashed arrow represents the reflected acoustic wave from the metal.}
\end{figure}

In experiments \cite{KOR1} there is a stationary X-ray emission in the $keV$ regime. It occurs from the surface of the metallic plate, in Fig.~\ref{fig1}, acted by shock waves. The emission spectra 
are continuous within the $keV$ scale with no narrow peaks of particular frequencies. 

Usually X-rays are generated by a beam of electrons which are accelerated in vacuum by an electric field of a few tens of $keV$ towards a metal target. Due to braking of electrons by the target,
Bremsstrahlung occurs with a continuous spectrum. In addition to that, characteristic radiation is possible. It corresponds to narrow peaks in the emission spectrum related to electron transition
among discrete levels of target atoms.  
 
At first sight, one can also expect that under shock waves a type of the emission radiation is either Bremsstrahlung or characteristic one. Below this situation is analyzed.

At the macroscopic description there is a discontinuity of medium velocity at the shock wave front and relatively long release wave behind the shock front. In reality the shock front cannot be of 
zero width. In the propagation of a shock through a real metal each atom processes a bounce when the shock wave reaches it. The atom returns to the initial position after the shock passed. Due to 
the atomic interaction, each atom is acted by the approaching shock front before it reaches the atom. So a width of the shock front is not shorter than distance between lattice sites. Therefore 
the perturbation of each atom, caused by the shock front, is a smooth function of time. The duration of this perturbation is of the characteristic time
\begin{equation}
\label{6} 
\Delta t\simeq\frac{{\rm width~of~shock~wave~front}}{{\rm shock~wave~speed}}\sim 10^{-13}s.
\end{equation}
The nominator in (\ref{6}) is of a few Angstroms and the denominator is $\sim 10^{5}cm/s$. 

The perturbation of each atom by the shock front, during the time $\Delta t$, is extremely adiabatic compared to the characteristic time $\hbar/(1keV)\sim 10^{-18}s$ of the $keV$ radiation. The 
fraction of $keV/\hbar$ frequencies in that perturbation is formally $\exp(-A)$, where $A\sim 10^{5}$ is proportional to ratio of the two time intervals. In \cite{KOR2} the atom perturbation by 
the shock wave was erroneously supposed to be a sudden ``jolt'' shorter than $10^{-18}s$.

When a perturbation of a lattice atom is not a shock but the usual sound wave, its characteristic time is no shorter than (\ref{6}). In this case the maximal wave vector is of the inverse 
inter-atomic distance and $\Delta t$ is inversely proportional to the Debye frequency.

Therefore the acoustic perturbation of electrons by a shock or usual sound is not able to excite atomic levels in the $keV$ region to produce characteristic radiation of that energy. Bremsstrahlung 
of $keV$ energy is also impossible because electrons in a solid are adiabatically dragged by the shock wave with velocity $\sim 10^{5}cm/s$. 

In experiments \cite{KOR1} the metallic surface, acted by shock waves and emitted X-rays, was far from the liquid producing cavitation mediated shock waves. So the internal dynamics of bubbles in 
the cavitating liquid was not relevant for the X-ray emission. 

This constitutes the paradoxical phenomenon of $keV$ X-ray generation under acoustic perturbations because this is incompatible with known effects. 
\section{THE MECHANISM}
\label{mec}
As follows from experiments \cite{KOR1}, the certain X-ray sources exist on the metal surface. They cannot be in the volume due to the short attenuation length ($\sim 1\mu$) of $keV$ radiation in 
metals. Those sources are in action during the shock wave pulse (\ref{6}). 

These puzzling issues look in contradiction with any combination of known effects. The generated frequency ($10^{18}s^{-1}$) is five orders of magnitude higher than one of the 
excitation source ($10^{13}s^{-1}$) related to the acoustic perturbation. This likely implies that shock waves produce the certain states, other than atomic ones, with low lying energy levels. 
Transitions to these levels result in the observed $keV$ X-ray generation. 

In principle, nuclei at lattice sites provide deeply lying states of the nuclear $MeV$ scale. But nuclear processes cannot be initiated by shock waves which are relatively soft. Low energy nuclear 
reactions are impossible due to the negligible probability of penetration through nuclear Coulomb barriers. 

Electron states in Coulomb fields are usually related to spatial scales generic with the Bohr radius. It seems to exclude a possibility of electron states in the Coulomb field other than atomic 
ones. However there is the short spatial scale (shorter than atomic one) related to electrons. It associates with the Lamb shift of electron energy levels and is clarified below. 

Suppose the electron to be in a Coulomb potential well in the ground state. Under the action of electromagnetic fluctuations the electron ``vibrates'' within the region (see 
\cite{WEL,MIGDAL,KOL,IVLEV3})
\begin{equation}
\label{1}
r_T=r_c\sqrt{\frac{4e^2}{\pi\hbar c}\ln\frac{\hbar c}{e^2}}\simeq 0.82\times 10^{-11}cm\,,
\end{equation}
where $r_c=\hbar/mc\simeq 3.86\times 10^{-11}cm$ is the electron Compton length. Due to ``vibrations'' the electron probes various parts of the potential well and therefore changes its energy by
the Lamb shift \cite{LANDAU}. 

Usually the quantum mechanical uncertainty of the wave function is much larger that $r_T$ and hence the Lamb shift is small according to weakness of the electron-photon interaction. Nevertheless
the opposite situation, when the spatial electron distribution is sharp, is possible. This is analyzed below.

The Schr\"{o}dinger equation has a singular solution which is $\psi\sim 1/R$ at small non-zero $R$. This reminds the electrostatic potential of a point charge. That solution does not exist
even formally since it requires the artificial singularity source $\delta(\vec R)$ in the right hand side of the Schr\"{o}dinger equation. After inclusion of fluctuating fields  $\vec u$ 
(electromagnetic, etc.) into the Schr\"{o}dinger formalism the averaged source $\langle\delta(\vec R-\vec u)\rangle$ gets spread out providing a non-existing source at a finite region of space. 

The singular solution can be supported not only by the artificial $\delta(\vec R)$. The electron mass is determined by the mean value of the Higgs field. This value, according to the Higgs 
formalism, weakly depends on electron distribution in space. However, when the electron distribution is singular, it can strongly disturb the Higgs field and the electron mass also becomes singular. 
This singularity of the electron mass serves as a natural source for the singular electron distribution. This contrasts with the above artificial $\delta(\vec R)$ source. 

That scenario is valid when all fluctuating fields are ``switched off''. Therefore although those singular forms are non-physical, nevertheless they are formal solutions in the absence of 
fluctuations. When fluctuations are ``switched on'' the singularity is washed out over the certain distance and the state becomes physical. 

The formation of those states cannot be described solely by QED since the mechanism of electron mass generation is involved. Details are given in Sec.e\ref{anon}.
\section{ANOMALOUS ATOMS}
\label{anon}
To follow a connection between a singularity of the electron density and associated properties of the electron mass we start with the mechanism of electron mass generation.
\subsection{Generation of electron mass}
\label{anomA}
In the Standard Model masses of electrons, other leptons, $W^{\pm}$ and $Z$ weak bosons, and quarks are generated by Higgs mechanism which involves the scalar Higgs field \cite{ENGL,HIGG,GUR}. 
Electron, as a fermion, acquires its mass by the connection between the fermion field $\psi$ and the Higgs field $\phi$. The Lagrangian
\begin{equation}
\label{20}
L=i\hbar c\bar{\psi}\gamma^{\mu}\tilde{D}_{\mu}\psi-G\bar{\psi}\phi\psi+L_{H}(\phi)+L_{g}
\end{equation}
contains the Higgs part
\begin{equation}
\label{21}
L_{H}(\phi)=\frac{1}{\hbar c}(D_{\mu})^{+}D^{\mu}\psi+\frac{1}{(\hbar c)^3}\left[\mu^2c^4\phi^{+}\phi-\lambda(\phi^+\phi)^2\right]
\end{equation}
and the gauge part $L_g$ that, for pure electromagnetic field, would be $-F^{\mu\nu}F_{\mu\nu}/16\pi$ where $F_{\mu\nu}=\partial_{\mu}A_{\nu}-\partial_{\nu}A_{\mu}$. The Yukawa term, depending on
the coupling $G$, is written in (\ref{21}) in a schematic form. The covariant derivatives $\tilde{D}_{\mu}$ and $D_{\mu}$ contain, in addition to partial derivatives $\partial_{mu}$, the parts
depending on gauge fields $W^{\pm}_{\mu}$, $z_{mu}$, and $A_{mu}$. In (\ref{20}) $\gamma^{\mu}$ are the Dirac matrices.

In our case the main contribution to the fermionic fields $\psi$ comes from the electron part. The isospinor $\phi=(0,v+h)$, besides the expectation value $v$, contains the fluctuation part $h$ 
with zero expectation value. The physical electron mass $m_0=Gv_0/c^2$ appears (in the Yukawa term) due to the finite expectation value $v_0=\mu c^2$ that relates to the ground state of $L_H$
\cite{ENGL,HIGG,GUR}. So the parameter $G=m/\mu$, where $\mu\sim 100GeV/c^2$, is the mass of the Higgs boson. One can estimate $G\sim 10^{-5}$. We normalize the Higgs field to have $\lambda=1/2$. 

As the first step, we consider the problem without fields of vector bosons $W^{\pm}_{\mu}$, $z_{mu}$, $A_{mu}$, and the Higgs field $h$. They can be included, as the second step, as given functions 
of space-time with the subsequent average on them. Analogous average on the photon field is performed in quantum electrodynamics. 

As follows from (\ref{21}), 
\begin{equation}
\label{22}
\nabla^2v+\frac{1}{\hbar^2c^2}\left(\mu^2c^4v-v^3\right)=\frac{\hbar c}{2}G\bar{\psi}\psi,
\end{equation}
where the right-hand side can be calculated according to Dirac quantum mechanics. One can put
\begin{equation}
\label{23}
v=\frac{c^2}{G}m,
\end{equation}
where the electron mass $m=m_0+\delta m(\vec R)$ in space according to variations of $v$. 

The electron spinors $\varphi$ and $\chi$, which form the total bispinor $\psi=(\varphi, \chi)$, satisfy the Dirac equations
\begin{eqnarray}
\nonumber
\left[\varepsilon-U(\vec R)+i\hbar c\vec\sigma\nabla\right]\varphi=mc^2\chi\\
\label{24}
\left[\varepsilon-U(\vec R)-i\hbar c\vec\sigma\nabla\right]\chi=mc^2\varphi.
\end{eqnarray}
Here $\varepsilon$ is the total relativistic energy, $\vec\sigma$ are Pauli matrices, and $U(\vec R)$ is some macroscopic potential. It follows from Eq.~(\ref{24}) that
\begin{equation}
\label{25}
\Theta=-\frac{i\hbar c\vec\sigma\nabla\Phi}{\varepsilon-U+mc^2}\,,
\end{equation}
where $\Phi=(\varphi+\chi)/\sqrt{2}$ and $\Theta=(\varphi-\chi)/\sqrt{2}$. The spinor $\Phi$ satisfies the equation 
\begin{equation}
\label{26}
-\nabla^2\Phi+\frac{\nabla\beta}{1+\beta}\left(\nabla\Phi-i\vec\sigma\times\nabla\Phi\right)+\frac{m^2c^2}{\hbar^2}\Phi=\frac{(\varepsilon-U)^2\Phi}{\hbar^2c^2}
\end{equation}
with the definition
\begin{equation}
\label{27}
\beta=\frac{c^2\delta m-U(\vec R)}{\varepsilon+m_0c^2}\,.
\end{equation}
Since the Dirac conjugate $\bar\psi=\psi^*\gamma^0$,
\begin{equation}
\label{28}
\bar\psi\psi=\varphi^*\chi+\chi^*\varphi=|\Phi|^2-|\Theta|^2.
\end{equation}
The electron density is
\begin{equation}
\label{29}
n=|\Phi|^2+|\Theta|^2.
\end{equation}
Below we consider the isotropic case when all values (including $U(R)$) depend on $R$. Under this condition the $i\vec\sigma$ term in (\ref{26}) disappears. To be specific one can put 
\begin{equation}
\label{30}
\Phi=\frac{1}{\sqrt{2}}{1\choose 1}F\,.
\end{equation}

When the deviation $\delta v$ from its equilibrium value $\mu c^2$ is small, it follows for $\delta m/m_0=\delta v/\mu c^2$
\begin{eqnarray}
\label{31}
&&\left(\nabla^2-\frac{2}{R^{2}_{c}}\right)\frac{\delta m}{m_0}\\
\nonumber
&&=\frac{G^2r_c}{2}\left[F^2-\frac{1}{(1+\varepsilon/m_0c^2)^2}\left(\frac{r_c\nabla F}{1+\beta}\right)^2\right],
\end{eqnarray}
where $r_c=\hbar/m_0c\simeq 3.86\times 10^{-11}cm$ is the electron Compton length and $R_c=\hbar/\mu c\sim 10^{-16}cm$ is the Compton length of the Higgs boson.

The electron density (\ref{29}) now reads
\begin{equation}
\label{32}
n=F^2+\frac{1}{(1+\varepsilon/m_0c^2)^2}\left(\frac{r_c\nabla F}{1+\beta}\right)^2.
\end{equation}
The equation for $F$ follows from (\ref{26})
\begin{equation}
\label{33}
-\nabla^2F+\frac{\nabla\beta}{1+\beta}\nabla F+\frac{F}{r^{2}_{c}}=\frac{(\varepsilon-U)^2F}{\hbar^2c^2}\,,
\end{equation}
where the mass variation in the term $1/r^{2}_{c}$ is not important.
\subsection{Singular solution in the absence of fluctuations}
\label{anomB}
If the electron is acted by a potential which is $m\Omega^2R^2/2$ at small $R$, the fluctuation radius (\ref{1}) contains $\ln\sqrt{mc^2/\hbar\Omega}$. So for a free electron $r_T=\infty$. When
the electron is in the attractive Coulomb potential, $\hbar\Omega$ is substituted by the Rydberg energy and we get the result (\ref{1}) \cite{WEL,MIGDAL}. Below we consider this case when the 
electron moves in the potential 
\begin{equation}
\label{34}
U(R)=-\frac{Ze^2}{\sqrt{R^2+r^{2}_{N}}}\,,
\end{equation}
where $r_N\sim 10^{-13}cm$ is the nuclear radius. 

At $R\sim r_c$ one can neglect $\nabla\beta$ term and $U$ in the right-hand side of (\ref{33}). In this case the solution of Eq.~(\ref{33}) takes the form
\begin{equation}
\label{35}
F=\frac{C}{R\sqrt{r_c}}\exp\left(-\frac{R}{\hbar c}\sqrt{m^{2}_{0}c^4-\varepsilon^2}\right),
\end{equation}
where $C$ is a dimensionless constant. We suppose $\varepsilon<m_0c^2$.

The function $\beta\sim Zr_N/\sqrt{R^2+r^{2}_{N}}$ because the Thompson radius $e^2/m_0c^2\sim 10^{-13}cm$ is on the order of $r_N$. At $R<r_c$ the main contribution is $F=C/R\sqrt{r_c}$\,. A 
correction to this result comes from the term $\nabla\beta\nabla F$ (rather than from $F/r^{2}_{c}$ term) in (\ref{33}) under the condition $|\nabla\beta|>R/r^{2}_{c}$ which is 
$R<(r_Nr^{2}_{c})^{1/3}$. We are restricted by a not large $Z$. With that condition the gradient terms in (\ref{33}) dominate resulting in the form
\begin{equation}
\label{36}
\frac{\partial F}{\partial R}=-C\,\frac{1+\beta(R)}{R^2\sqrt{r_c}}\,,\hspace{0.5cm}R<(r_Nr^{2}_{c})^{1/3}.
\end{equation}

Under the additional condition $R_c<R$ the gradient term in the left-hand side of Eq.~(\ref{31}) is small. But in the right-hand sides of Eqs.~(\ref{31}) and (\ref{32}) the gradient terms dominate.
This results in the mass correction
\begin{equation}
\label{37}
\frac{\delta m(R)}{m_0}=\frac{G^2}{4}r_cR^{2}_{c}n(R),\hspace{0.5cm}R_c<R<r_c,
\end{equation}
where the electron density
\begin{equation}
\label{38}
n(R)=\frac{C^2}{(1+\varepsilon/m_0c^2)^2}\frac{r_c}{R^4}\,,\hspace{0.5cm}R_c<R<r_c.
\end{equation}

The contribution to $\nabla\beta$ from the $\delta m$ term in (\ref{27}) is principal at $R<(r_NR^{2}_{c})^{1/3}$. This provides the singular contribution of the $\nabla\beta$ term in 
Eq.~(\ref{33}). From Eqs.~(\ref{38}) and (\ref{37}) we see how the singularity in the electron distribution, according to Sec.~\ref{mec}, is connected with the singularity of the electron mass in 
the formal absence of fluctuations. 

At distances $R$ shorter than $R_c$ the correction $\delta m/m_0$ becomes large and the left-hand side of the equation (\ref{31}), based on the expansion around the equilibrium 
value $\mu c^2$ of $v$, is not correct. However since the right-hand side of (\ref{31}) remains singular, this leads to a singular $v$. In turn, the $\nabla\beta$ term in Eq.~(\ref{33}) remains 
singular serving as a singularity source for $F$. 
\subsection{Anomalous atoms}
\label{anomC}
So the singular solution for $F$, if to artificially ``switch off'' fluctuations, formally exists. As the second step, one has to average this solution on fluctuating fields of 
$W^{\pm}_{\mu}$, $Z_{\mu}$, $A_{\mu}$, and the Higgs field $h$. These fluctuations wash out the singularity on a finite distance. The main contribution to this effect comes from the massless photon 
field $A_{\mu}$. The corresponding fluctuation length is (\ref{1}). Massive fields of other gauge bosons and $h$ relate to a shorter fluctuation length. The resulting state becomes physical and 
strongly localized within the sphere of the radius $r_T$. Due to the condition $R_c<r_T<r_c$ the form (\ref{38}) is to be averaged. One can approximate the averaged electron density as 
\begin{equation}
\label{39}
n(R)=\frac{r_T}{\pi^2(R^2+r^{2}_{T})^2}\,,
\end{equation}
which accounts for the normalization condition $\int nd^3R=1$ by a proper choice of the constant $C$ in (\ref{38}). The mass correction at the region $R\sim r_T$ can be estimated as 
$\delta m/m_0\sim G^4\sim 10^{-20}$. 

Since the electron is localized at the region $R<r_T$ its kinetic energy is enhanced. The enhancement of the kinetic energy $\hbar c/r_{T}\sim 1MeV$ at that region is compensated by the reduction 
of the zero point photon energy at the same region (anomalous well) \cite{IVLEV3,IVLEV4}
\begin{figure}
\includegraphics[width=5.5cm]{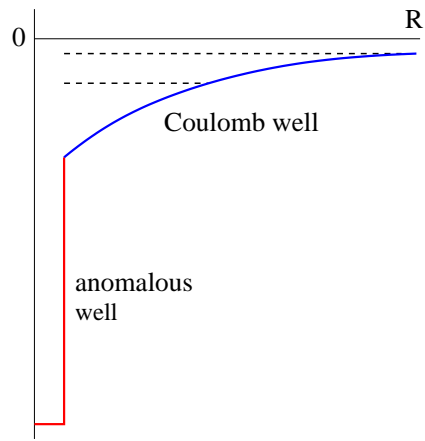}
\caption{\label{fig2}Scheme of anomalous atom. The usual Coulomb well goes over into the narrow ($\sim 10^{-11}cm$) and deep ($\sim 1MeV$) anomalous well where the electron spectrum is continuous. 
Dashed horizontal lines represent energy levels in the initial Coulomb field. The typical distance between these levels is in the $keV$ region.}
\end{figure}
\begin{equation}
\label{40}
\sum\frac{\hbar\omega}{2}-\left(\sum\frac{\hbar\omega}{2}\right)_0.
\end{equation}
Here the last term corresponds to absence of the electron. The first term is spatially dependent through the variable density of states \cite{IVLEV3,IVLEV4}. As a result, the energy (\ref{40}) 
corresponds to the narrow ($\sim 10^{-11}cm$) and deep ($\sim 1MeV$) well. Analogous well is formed in the Casimir effect of attraction of two atoms when, in contrast, the well is shallow and wide. 

We see that the electron is localized within the sphere $R<r_T$. It is acted, besides the Coulomb field of the bare nucleus, by the narrow and deep (anomalous) well. This configuration can be 
treated as {\it anomalous atom}. The total potential in anomalous atom is shown in Fig.~\ref{fig2}. 

In the usual potential well in quantum mechanics energy levels are quantized due to absence of a singularity inside the well. In our case such a condition does not exist due to smearing of the 
singularity by fluctuating fields. For this reason, the energy spectrum in the anomalous well is continuous. 

The anomalous state is an exact electron-photon one where $e^2/\hbar c$ is accounted for exactly. Therefore the lifetime of each state from the
continuous set is infinite. This is analogous to the exact electron-photons state with the continuous spectrum of infinite lifetime states studied in Ref.~\cite{IVLEV2}. 

The conclusion about infinite lifetime of states in anomalous well is valid when they are formed in a static potential whose minimum has a fixed position in space. In a crystal lattice a minimum 
position is determined by positions of lattice sites. Due to their thermal vibrations electrons in anomalous well also vibrates resulting in Bremsstrahlung and hence to a very small but finite 
lifetime of states in anomalous well \cite{IVLEV3}.

We emphasize that the concept of anomalous atoms cannot be formulated within QED only since the mechanism of electron mass generation is involved. 
\subsection{Creation of anomalous atoms}
\label{anomD}
The anomalous electron state has the typical spatial scale of $r_T\sim 10^{-11}cm$ corresponding to strong oscillations in space. The typical spatial scale of electron in condensed matter is
$10^3r_T$. Therefore the matrix element between these states is of the type $\exp(-1000)$. In contrast, when a perturbation is of a short scale, comparable with $r_T$, the probability of anomalous 
state creation is not small. 

Each lattice site in a solid, due to zero point oscillations in the crystal field, is smeared out within the Debye-Waller distance of $(0.01-0.1)\AA$ \cite{ZIM}. In experiments \cite{KOR1} under 
the action of the shock wave, reflected from the surface $A$ in Fig.~\ref{fig1}, a standing de Broglie wave of lattice sites $\cos(2MvR/\hbar)$ is formed. Here $M$ is the mass of the lattice site 
and $v$ is the shock wave velocity. This provides the spatially varied charge density with the typical scale
\begin{equation}
\label{41}
R=\frac{\hbar}{2Mv}\,.
\end{equation}

For the steel plate in experiments \cite{KOR1} $M\simeq 0.93\times 10^{-22}g$ and the sound velocity is $s\simeq 5.93\times 10^{5}cm/s$. The velocity $v$ exceeds $s$ in the shock wave. One can
estimate the typical distance (\ref{41}) for the iron case in the form $R/r_T\simeq 1.16s/v$. In experiments \cite{KOR1} cavitation mediated shock waves were involved. These waves are slow that is
their velocity $v$ slightly exceeds the speed of sound $s$ \cite{GOM}. For iron atoms the condition $R/r_T\sim 1$ of anomalous well formation holds. Note that approximately $s\sim 1/\sqrt{M}$.
So by the action of cavitation mediated (slow) shock waves anomalous states can be formed solely around heavy atoms like iron. In experiments \cite{KOR2} it was copper based anomalous states.

For not heavy atoms (like oxygen) slow shocks cannot create anomalous states. When a shock wave is reflected in the water or in a biological medium, standing de Broglie wave of oxygen atoms 
corresponds to the spatial charge variation (\ref{41}) on the scale $R/r_T\simeq 16s/v$. For this case $M\simeq 2.65\times 10^{-23}g$ and $s\simeq 1.5\times 10^{5}cm/s$. The above interference 
phenomenon is expected with counter shock waves. We see that to generate anomalous atoms in a biological medium (the condition $R\sim r_T$) the shock wave should be fast, $v\sim (10-15)s$. Such 
fast shock treatment of biological objects is expected to result in accumulation of anomalous atoms which are not chemically active and may drastically change the biological medium.  

In experiments \cite{KOR1}, where shocks are slow, iron atoms play a substantial role for X-ray emission. Without the steel plate in Fig.~\ref{fig1} X-rays are not registered by the X-ray film. In 
this case the metallic rode, which also emits X-rays, is too far from the film. 
\subsubsection{Emission of high energy quanta}
Before the reflection of the shock wave from the metal surface usual electron states exist in the Coulomb potential of the lattice site shown by dashed lines in Fig.~\ref{fig2}. We do not consider 
conduction electrons. Under the interference of incident and reflected de Broglie waves of lattice sites the anomalous well is formed during the characteristic time $\hbar/1MeV\sim 10^{-21}s$. 

That ``jolt'' essentially violates the systematics of electron states. First, emission of $MeV$ quanta is expected. Second, the old stationary states (dashed lines in Fig.~\ref{fig2}) become 
non-stationary characterized by the flux toward the anomalous well. The distance between old levels is in the $keV$ region. Therefore this process is characterized by the time 
$\hbar/1keV\sim 10^{-18}s$ and by emission of quanta of the continuous spectrum in the region of $keV$. This X-ray emission was observed in experiments \cite{KOR1}. 
\subsubsection{Binding energy of anomalous atom}
The electrons have a tendency to go down in energy in the anomalous well in Fig.~\ref{fig2}. It is energetically favorable to acquire other electrons with the energy gain $\hbar c/r_T$ per each.
The total energy gain is estimated as 
\begin{equation}
\label{42}
\Delta E\simeq -N\left(\frac{\hbar c}{r_T}+\frac{Ze^2}{r_T}\right)+\frac{N^2e^2}{2r_T}\,,
\end{equation}
where $N$ is the number of acquired electrons. The second term is the Coulomb interaction with the nucleus of the charge $Ze$. The third term is due to the Coulomb repulsion of acquired electrons. 
The maximal energy gain corresponds to the maximal $N$ which cannot be larger than $Z$. Otherwise the confining potential, providing the finite $r_T$, disappears. Putting $N=Z$, one obtains the
total binding energy of the anomalous atom
\begin{equation}
\label{43}
\Delta E\simeq -Z\frac{\hbar c}{r_T}\left(1+\frac{Ze^2}{\hbar c}\right).
\end{equation}
The size $10^{-11}cm$ of the anomalous atom is one thousand times less than one of a usual atom. The energy $\hbar c/r_T\simeq 2.4MeV$. For iron $Z=26$ and therefore the binding energy of the iron 
based anomalous atom is $\Delta E\simeq -74MeV$. 
\subsection{Summary of features of anomalous atoms}
\label{anomE}
(1) Anomalous atom is based on the usual atomic nucleus with the charge $Ze$. 

(2) Due to a spatial redistribution of the electromagnetic zero point energy the spherical well of the radius $\sim 10^{-11}cm$ is formed around the nucleus. The depth of this anomalous well is 
approximately a few $MeV$. 

(3) $Z$ electrons are expected to get acquired by the anomalous well releasing their total energy of about $2.4Z\,MeV$. This is the binding energy of anomalous atom.

(4) The size of anomalous atom $10^{-11}cm$ is one thousand times less than hydrogen atom.
\section{NEUTRON EMISSION BY SHOCK WAVES}
In this section neutron emission by acoustic perturbation is analyzed.
\subsection{Neutrons from liquids}
\label{neutA}
In experiments \cite{CARD1,CARD2} neutron emission was registered from cavitating (by the acoustic source) water solution of iron salts. It was no residual radioactive isotops in the matter since 
neutron emission appeared gradually upon increasing of acoustic power of the generator producing cavitation. It was a matter of special care to distinguish between detection of $\gamma$-quanta
and neutrons. In particular, the neutron detectors used did not react on $\gamma$-quanta from the tested $^{60}{\rm Co}$ sample. 

Bubbles, created by the acoustic source, emit cavitation mediated shock waves \cite{GOM}. But before this emission the bubble dynamics is complicated. Each bubble, of the micron size, rapidly 
shrinks as in the phenomenon of sonoluminescence \cite{PUTT,BRE}. 

During that implosion the energy is accumulated inside the bubble resulting in the UV emission up to $10eV$ in quanta energy. This energy is much smaller than the nuclear $MeV$ scale. Therefore 
the neutron source in \cite{CARD1,CARD2} is not inside micron size bubbles.

In experiments \cite{CARD1,CARD2}, as wel as in \cite{KOR1}, cavitation mediated shock waves perturbed atoms of the liquid. With respect to nuclear processes (time interval 
$\hbar/1MeV\sim 10^{-21}s$) this perturbation is extremely adiabatic (time interval $10^{-13}s$) and cannot directly lead to neutron emission. However, when those shock waves reflect from the
liquid-solid border, anomalous atoms are formed around iron nuclei (Sec.~\ref{anomD}). During this formation emission of $MeV$ quanta occurs that can result in high energy (nuclear) processes. 
As shown in Sec.~\ref{anomD}, the total energy release in formation of the iron based anomalous atom can be larger than $70MeV$.

In experiments \cite{CARD1,CARD2} on neutron emission from cavitating liquid the presence of iron atoms was essential. Without iron salts neutron emission was absent. Analogously in \cite{KOR1} 
it was no X-ray emission without iron parts. In the both cases shock waves were slow (cavitation mediated) and solely heavy ions could result in formation of anomalous states (Sec.~\ref{anomD}).
 
So neutron emission in experiments \cite{CARD1,CARD2} is expected to occur from the liquid-solid border in formation of anomalous atoms by cavitation mediated shock waves.  
\subsection{Neutrons from solids}
\label{neutB}
In Refs.~\cite{CARD3,CARP1,CARP2} neutron emission from solids was reported. The driving force for that was shock waves generated by fractures of microdefects in iron-reach natural rocks. 
Generation and propagation of shock waves in elastoplastic solids is described on the level of solid state physics \cite{DUR}. These processes are not related to formation of high-density fluid, 
plasma, etc. involving $MeV$ energies. Therefore in this case there is also the paradoxical mismatch between the adiabatic perturbation ($10^{-13}s$) and the fast response ($10^{-21}s$) related 
to neutron emission. 

This contradiction disappears if to account for formation of anomalous states on the border of the solid as in Sec.~\ref{anomD}. Those states relate to deep ($MeV$ scale) wells for electrons which, 
falling down in the well, emit high energy quanta. These quanta participate in usual nuclear reactions producing neutrons. So neutron emission under acoustic perturbation in solids is also expected 
from the border of the sample. 
\section{DISCUSSIONS}
\label{disc}
Under propagation of a shock wave in a condensed matter each atom processes a bounce returning to its initial position after the shock passed. The most rapid process in the atomic motion relates 
to the shock front. After that atoms move relatively slow in the release wave to get back to the initial positions. The shock front cannot be infinitely narrow as at a macroscopic description 
when the medium is supposed to be continuous. In reality the inter-atomic distance is finite leading to the finite front width. As a result, under the shock wave the atomic displacement becomes 
smooth, as a function of $t$, with the characteristic time $\Delta t\sim 10^{-13}s$. 

On the other hand, for $keV$ X-rays the typical time $\hbar/1keV\sim 10^{-18}s$ is five orders of magnitude shorter than the exciting shock pulse. So the smooth atomic displacement is extremely
adiabatic with respect to emitted X-rays. Despite the atomic displacement contains all frequencies around the main one, $1/\Delta t$, the fraction of frequencies $10^{18}s^{-1}$ is exponentially
small, $\exp(-A)$, where $A\sim 10^5$ is proportional to the ratio of two time intervals.  

Therefore the perturbation of electrons by the shock wave (if to account for usual mechanisms) cannot result in Bremsstrahlung of $keV$ energy. Also that perturbation is not able to excite atomic 
levels in the $keV$ region to produce a characteristic radiation of that energy. In experiments \cite{KOR1} the metallic surface, acted by shock waves and emitted X-rays, was far from the liquid 
producing cavitation mediated shock waves. So the internal dynamics of bubbles in the cavitating liquid was not relevant for the X-ray emission. Nevertheless in experiments \cite{KOR1,HAG} X-ray 
emission was registered. 

The resolution of this contradiction is in formation of deep ($\sim 1MeV$) and narrow ($\sim 10^{-11}cm$) anomalous wells localized in the vicinity of a nucleus. Due to transitions to this well 
from atomic states electrons emit $keV$ quanta. Anomalous atoms correspond to the ground state. But transitions to this state of usual electrons in condensed matter requires a perturbation which 
rapidly (on the scale $10^{-11}cm$) varies in space. Therefore for usual perturbations, with the scale $10^{-8}cm$, this transition probability is of the type $\exp(-1000)$. 

When a shock wave in a solid reflects from its border, the length of the standing de Broglie wave is short providing a rapidly varying charge density. Under this perturbation the anomalous state 
is formed during the time interval $\hbar/1MeV\sim 10^{-21}s$. This results in the observed emission of $keV$ X-rays. In that process quanta of $MeV$ scale also can be emitted. As a result, $Z$ 
electrons fill the anomalous well formed around the nucleus with $Ze$ charge. Shock waves collide the border approximately every $10^{-4}s$ which is the repetition time of the process.  

In experiments \cite{CARD1,CARD2,CARD3,CARP1,CARP2} neutron emission was registered from liquids and solids acted by shock waves. There is the common feature of those experiments and \cite{KOR1}: 
the paradoxical mismatch between the adiabatic ($10^{-13}s$) perturbation by the shock wave and the short time response. This response corresponds to $10^{-18}s$ for the $keV$ X-ray emission and 
$10^{-21}s$ for nuclear processes (neutron emission). Within usual mechanisms, the probability of these processes, exponentially depending on the ratio of two time intervals, is zero in reality. 

It happens that neutrons, observed in experiments \cite{CARD1,CARD2,CARD3,CARP1,CARP2}, are emitted in usual nuclear reaction initiated by high energy quanta. Low energy nuclear reaction are 
impossible due to high Coulomb barriers to be passed by charged nuclei. Those quanta come from electron transitions in deep ($MeV$ scale) anomalous wells. The well is formed by the local reduction 
of zero point electromagnetic energy. Similar reduction, on much larger spatial scale, occurs in the Casimir effect. The formation of anomalous atoms cannot be described solely by QED since the 
mechanism of electron mass generation is involved. It is unusual that the electron subsystem in condensed matter relates to $MeV$ energies which are typical for nuclear processes. 

~~~~~~

\section{CONCLUSIONS}
At first sight, experimentally observed $keV$ X-ray and neutron emissions by acoustic perturbations of liquids and solids look paradoxical. A very adiabatic perturbation can provide solely an
exponentially small contribution to high energy effects. Substantially high energies in the problem cannot be of a nuclear origin because low energy nuclear reactions are impossible due to high 
Coulomb barriers to be passed by charged nuclei. 

High energies unexpectedly come from the different source which is of the electron origin. This is the deep ($\sim 1MeV$) and narrow ($10^{-11}cm$) anomalous wells for electrons created by the 
local reduction of electromagnetic zero point energy. The formation of anomalous states cannot be described solely by quantum electrodynamics since the mechanism of electron mass generation is 
involved. 

Anomalous wells correspond to the ground state but formation of them requires a rapid variation in space (on the distance $10^{-11}cm$) of the charge density. This condition holds under shock
waves reflections. 
\acknowledgments
I thank J. L. Diaz Cruz, J. Engelfried, P. L. Hagelstein, J. Knight, E. B. Kolomeisky, A. A. Kornilova, M.~N.~Kunchur, A. M. Loske, and N. N. Nikolaev for discussions and remarks. This work was 
supported by CONACYT through grant number~237439.


\begin{thebibliography}{8}

\bibitem{KOR1}

A. A. Kornilova, V. I. Vysotskii, N. N. Sysoev, N. K. Litvin, V. I. Tomak, and A. A. Bazov, J. Surf. Invest. X-ray, Synchrotron, and Neutron Tech., {\bf 4}, 1008 (2010).

\bibitem{KOR2}

A. A. Kornilova, V. I. Vysotskii, N. N. Sysoev, and A. V. Desyatov, J. Surf. Invest. X-ray, Synchrotron, and Neutron Tech., {\bf 2}, 275 (2009).

\bibitem{HAG}

P. Hagelstein, private communication (2017).

\bibitem{CARD1}

F. Cardone, G. Cherubini, R. Mignani, W. Perconti, A. Petrucci, F. Rosetto, and G. Spera, arXiv:0710.5115.

\bibitem{CARD2}

F. Cardone, G. Cherubini, R. Mignani, and A. Petrucci, arXiv:0812.1272. 

\bibitem{PUTT}

S. J. Putterman and K. R. Weniger, Annu. Rev. Fluid Mech. {\bf 32}, 445 (2000).

\bibitem{BRE}

M. P. Brenner, S. Hilgenfeldt, and D. Lohse, Rev. Mod. Phys. {\bf 74}, 425 (2002).

\bibitem{GOM}

R. Pecha and B. Gompf, Phys. Rev. Lett. {\bf 84}, 1328 (2000).

\bibitem{BREN}

C. E. Brennen, {\it Cavitation and Bubble Dynamics} (Oxford University Press, 1995).

\bibitem{VOG}

A. Vogel, S. Busch, and U. Parlitz, J. Acoust. Soc. Am. {\bf 100}, 148 (1996).

\bibitem{CARD3}

F. Cardone, A. Carpinteri, and G. Lacidogna, Phys. Lett. A {\bf 373}, 4158 (2009).

\bibitem{CARP1}

A. Carpinteri, in {\it Acoustic, Electromagnetic, Neutron Emissions from Fracture and Earthquakes}, (Springer International Publishing, Switzerland, 2015).

\bibitem{CARP2}

A. Carpinteri, F. M. Cur\`{a}, R. Sesana, A. Manuello, O. Borla, and G. Lacidogna, in {\it Acoustic, Electromagnetic, Neutron Emissions from Fracture and Earthquakes}, (Springer International 
Publishing, Switzerland, 2015).

\bibitem{DUR}

T. Cohen and D. Durban, Proc. R. Soc. A {\bf 470}, 0061 (2014).

\bibitem{IVLEV3}

B. I. Ivlev, Can. J. Phys. {\bf 95}, 514 (2017); B. I. Ivlev, arXiv:1512.08504.

\bibitem{IVLEV4}

B. I. Ivlev, arXiv:1701.00520.

\bibitem{PRE}

P. Zhong, C. J. Chuong, and G. M. Preminger, J. Acoust. Soc. Am. {\bf 94}, 29 (1993).

\bibitem{WEL}

T. A. Welton, Phys. Rev. {\bf 74}, 1157 (1948).

\bibitem{MIGDAL}

A. B. Migdal, {\it Qualitative Methods in Quantum Theory} (Adison-Wesley, 2000).

\bibitem{KOL}

E. B. Kolomeisky, arXiv:1203.1260.

\bibitem{LANDAU}

V. B. Berestetskii, E. M. Lifshitz, and L. P. Pitaevskii, {\it Quantum Electrodynamics} (Pergamon, New York, 1980).

\bibitem{ENGL}

F. Englert and R. Brout, Phys. Rev. Lett. {\bf 13}, 321 (1964).

\bibitem{HIGG}

P. Higgs, Phys. Rev. Lett. {\bf 13}, 508 (1964).

\bibitem{GUR}

G. S. Guralnik, C. R. Hagen, and T. W. B. Kibble, Phys. Rev. Lett. {\bf 13}, 585 (1964).

\bibitem{IVLEV2}

B. I. Ivlev, Can. J. Phys. {\bf 94}, 1253 (2016); B. I. Ivlev, arXiv:1510.01279.

\bibitem{ZIM}

J. M. Ziman, {\it Principles of the Theory of Solids} (University Press, Cambridge, 1964).

\end{thebibliography}
\end{document}